\documentclass[aps,prd,showpacs,preprintnumbers]{revtex4}
\usepackage{graphics}
\def \beq{\begin{equation}}
\def \eeq{\end{equation}}
\def \beqa{\begin{eqnarray}}
\def \eeqa{\end{eqnarray}}
\newcommand{\cC}{{\cal C}}

\def \etal{{\sl et al.~\/}}
\def \ibid{{\sl ibid~\/}}
\def \arnp{{\sl Ann.\ Rev.\ Nucl.\ Part.\ Sci.~\/}}
\def \eup{{\sl Eur.\ Phys. ~\/}}
\def \jhep{{\sl J.\ H.\ E.\ P.~\/}}
\def \jpcs{{\sl J.\ Phys.\ Conf.\ Ser.~\/}}
\def \mpla{{\sl Mod.\ Phys.\ Lett.\ A~\/}}
\def \np{{\sl Nucl.\ Phys.~\/}}
\def \npsl{{\sl Nucl.\ Phys.\ Proc.\ Suppl.~\/}}
\def \pl{{\sl Phys.\ Lett.~\/}}
\def \pr{{\sl Phys.\ Rev.~\/}}
\def \prp{{\sl Phys.\ Rept.~\/}}
\def \prgnp{{\sl Prog.\ Part.\ Nucl.\ Phys.~\/}}
\def \prl{{\sl Phys.\ Rev.\ Lett.~\/}}
\def \rmp{{\sl Rev.\ Mod.\ Phys.~\/}}
\begin{document}
 
\title{Susceptibilities and speed of sound from PNJL model}
\author{Sanjay K.\ \surname{Ghosh}}
\email{sanjay@bosemain.boseinst.ac.in}
\author{Tamal K.\ \surname{Mukherjee}}
\email{tamal@bosemain.boseinst.ac.in}
\affiliation{Department of Physics, Bose Institute, 
          93/1, A.P.C. Road, Kolkata 700 009, India.}
\author{Munshi G.\ Mustafa}
\email{munshigolam.mustafa@saha.ac.in}
\author{Rajarshi \surname{Ray}}
\email{rajarshi.ray@saha.ac.in}
\affiliation{Theory Division, Saha Institute of Nuclear Physics, 
          1/AF, Bidhannagar, Kolkata 700 064, India.}

\begin{abstract}

We present the Taylor expansion coefficients of the pressure in 
quark number chemical potential $\mu_0=\mu_B / 3=\mu_u=\mu_d$, for the 
strongly interacting matter as described by the PNJL model for two light 
degenerate flavours of quarks $u$ and $d$. The expansion has been 
done upto eighth order in $\mu_0$, and the results are consistent
with recent estimates from Lattice. We have further obtained the
specific heat $C_V$, squared speed of sound $v_s^2$ and the conformal 
measure $\cC$.

\end{abstract}
\pacs{12.38.Aw, 12.38.Mh, 12.39.-x \hfill
SINP/TNP/06-04}
\maketitle

\section{Introduction}

Strongly interacting matter at non-zero baryon density and high
temperature is a subject of great interest for a wide spectrum of
physicists. A deep understanding of the different facets of strongly 
interacting matter, especially, the physics of colour deconfinement
might help us to get a better picture of various astrophysical and
cosmological phenomena. Huge accelerators have been built at CERN, 
Geneva \cite{lhc} and at RHIC, Brookhaven \cite{rhic} to collide 
beams of heavy nuclei at relativistic energies to recreate the matter 
in such extreme conditions. The focus of the heavy-ion physics 
program is to study and understand properties like the emergence of 
macroscopic collective phenomenon and their role on the evolution
of a strongly interacting system. The analysis of data so obtained 
requires a proper understanding of Quantum Chromodynamics (QCD), the 
theory of strong interactions. The thermodynamic properties are obtained
from the QCD equation of state (EOS) and various transport coefficients.

The weak coupling expansion of the free energy within perturbative
QCD (pQCD), is presently known \cite{pqcd} to order $\alpha_s^{5/2}$ or
$g^5$. However, in spite of the higher order, the series show a 
deceptively poor convergence except for coupling constants as low as 
$\alpha_s < 0.05$, corresponding to temperature as high as $10^5~T_c$.

There has been significant recent activity to improve on this
convergence by Hard Thermal Loop (HTL) re-summation schemes \cite{htl} 
which give a description of the plasma in terms of weakly interacting 
hard and soft quasi-particles \cite{qs1,qs2,qs3}. The former are massive
excitations with masses $m_D \sim gT$, while latter are either on-shell 
collective excitations or virtual quanta exchange in the interactions 
between hard particles. In contrast to ordinary perturbation theory, 
the HTL and its next-to-leading order approximations are well controlled
in the regime above $5T_c$, which is however, much higher than the 
temperature range in present day heavy-ion collisions.
In addition the different approaches within the same general framework
of HTL approximations lead to different predictions \cite{htl1,htl2,
htl3,htl4,htl5,htl6,htl7}, and cannot be made fully systematic when
applied to thermodynamics. At the same time, this approach lacks a 
reliable and comprehensive description below $5 T_c$ due to their 
stronger sensitivity to non-perturbative collective phenomena.

There is also a non-perturbative first principle method to compute
the contribution of the soft fields to the thermodynamics. At high
temperatures the hard modes get large temperature dependent masses,
which are then integrated out leaving a three-dimensional effective
theory. This method is known as dimensional reduction 
\cite{dmr0,dmr1,dmr2}. However, as expected, this method is not 
well-suited for lower temperature and the susceptibilities
calculated from these method \cite{dmr3} do not agree with those
of Lattice QCD computations below $\sim 2 T_c$.

Exploring the qualitative features of strongly interacting matter and 
making quantitative predictions of its properties from first principles
is the central goal of numerical studies of equilibrium thermodynamics of 
QCD within the framework of Lattice regularization. Over the years this
formulation has given us a wealth of information (for a review see 
Ref.\cite{laer}). It has now been established that there is only a 
crossover of normal hadronic matter to a state of deconfined quarks and 
gluons at temperature $T_c \sim 200 ~\rm{MeV}$. At this temperature,
deconfinement of {\sl color} charge and restoration of chiral symmetry
is found to occur simultaneously. Moreover, the equation 
of state, various susceptibilities and transport coefficients have been 
obtained. To perform Lattice QCD computations at non-zero chemical 
potential however is still a non-trivial task due to the complex fermion 
determinant. Recently though, there have been various new developments 
to tackle this problem for small chemical potentials 
\cite{smalmu1,smalmu2,smalmu3,smalmu4,smalmu5,eight,sixx}. 

Generally the QCD inspired phenomenological models are much easier to
handle compared to Lattice or the perturbative QCD calculations. 
But in all these models despite their simplicity, the absence of
a proper order parameter for deconfinement transition adds to the
uncertainties inherent in such studies and hence reduces the predictive
power of such models.

The thermal average of the Polyakov loop can be considered to be the 
order parameter for deconfinement transition \cite{polyl}. Hence a 
judicious use of the Polyakov loop in effective models may prove to be
of great advantage. Some important developments have taken place in 
obtaining effective theories for Polyakov loop from that of the temporal
background gauge field \cite{polyd1,polyd2,polyd3}. It is expected that
such an
effective theory for the Polyakov loop should work at least near the 
phase transition where long distance physics become important. More 
recently, the parameters in these effective theories have been fixed 
\cite{pisarski1,pisarski2} using the Lattice data (similar comparisons
of perturbative effects on Polyakov loop with Lattice data above the 
deconfinement transition was studied in \cite{salcd}). These 
Polyakov loop models have been applied in the context of cosmology 
\cite{ajit}.

For non-zero chemical potential there are various QCD inspired models 
which indicate (see e.g. Refs.\cite{qcdpd1,qcdpd2,qcdpd3,qcdpd4,qcdpd5})
that at low temperatures there is a possibility of first order phase 
transition for a large baryon chemical potential $\mu_{B_c}$. This 
$\mu_{B_c}$ is supposed to decrease with increasing temperature. Thus 
there is a first order phase transition line starting from 
($T=0$, $\mu_B=\mu_{B_c}$) on the $\mu_B$ axis in the ($T$,$\mu_B$) 
phase diagram which steadily bends towards the ($T=T_c$, $\mu_B=0$) 
point and may actually terminate at a critical end point (CEP)
characterized by ($T=T_E$, $\mu_B=\mu_{B_E}$), which can be detected via
enhanced critical fluctuations in heavy-ion reactions \cite{rajag1}.
The location of this CEP has become a topic of major importance in 
effective model studies (see e.g. Ref.\cite{njltype}). Also on the Lattice
the CEP was located for the physical \cite{fdkz1} and for somewhat larger 
\cite{fdkz2} quark masses using the reweighting technique of \cite{smalmu1},
and for Taylor expansion method in \cite{eight}. 

Thus the broad picture of QCD thermodynamics is becoming more 
transparent with the combination of research in the areas of 
perturbative QCD, effective models and Lattice computations.
In this paper we study some of the thermodynamic properties of strongly 
interacting matter using the Polyakov loop $+$ Nambu-Jona-Lasinio (PNJL)
model \cite{pnjl0}. The motivation behind the PNJL model is to couple
the chiral and deconfinement order parameters inside a single framework. 
Similar approach to understand QCD thermodynamics is being actively
pursued by various authors (see e.g. Ref.\cite{enrique} and references
therein). Here we shall first compute the EOS and the quark number 
susceptibilities. Susceptibilities in general, are related to 
fluctuations via the fluctuation-dissipation theorem. The difference in
fluctuations of various conserved quantities like baryon number, 
electric charge etc. in the hadronic and deconfined phases is supposed
to signal the phase transition between these two phases in heavy-ion 
reactions \cite{flucth}. Measurements of these fluctuations have taken 
a central place in the heavy-ion collisions \cite{flucex2}. 
Recently, computations on the Lattice have obtained many of these 
susceptibilities at zero chemical potential \cite{eight,sixx}, and our 
first aim would be to compare the corresponding quantities extracted 
from the PNJL model. As we shall show here that the agreement of the 
values obtained from the PNJL model is quite satisfactory with the 
Lattice data. 

Subsequently we have studied few more quantities of interest. One such 
quantity is the specific heat $C_V$, which is related to the 
event-by-event temperature fluctuations \cite{ebe1}, and mean transverse
momentum fluctuations \cite{ebe2} in heavy-ion reactions. These 
fluctuations should show a diverging behaviour near the CEP. Next, we 
obtain the speed of sound (basically its square, $v_s^2$), which 
determines the flow properties in heavy-ion reactions 
\cite{elp1,elp2,elp3,elp4}. Using the proper hydrodynamic equations 
including the speed of sound it is possible to analyze the rapidity 
distribution of secondary particles in collision experiments 
\cite{beda}. Finally we obtain the conformal measure 
$\cC=\Delta/\epsilon$, where $\Delta = \epsilon - 3P$ is the interaction
measure and $\epsilon$ and $P$ are respectively the energy density and 
pressure of strongly interacting matter. As has been pointed out in 
\cite{swa1,swa2}, the conformal measure seems to be emerging as an 
important measure to draw similarities between long distance physics of
QCD and conformal field theory, with results coming from both the areas
of AdS/CFT correspondence \cite{adscft}, RHIC data \cite{adsrhic} and 
Lattice computations \cite{adslat}.

The plan of the paper is as follows. In section 2, we discuss the
PNJL model briefly. In section 3, we present the formalisms. The EOS 
has been obtained from the Lagrangian of PNJL model in mean-field 
approximation. The Taylor expansion coefficients of pressure with 
respect to the quark number chemical potential $\mu_0$,
is written down. We also present the various formulae for $C_V$, $v_s$
and $\cC$. In section 4, we present our results and comparisons with
Lattice data. Finally we conclude with a discussion in section 5.

\section{PNJL Model}

The PNJL model was first introduced in Ref.\cite{pnjl0} to couple the
Nambu-Jona-Lasinio (NJL) model \cite{njl1} (see Ref.\cite{njls} for most 
recent developments), with the Polyakov loop. Recently, this has been 
extended in Refs.\cite{pnjl1,pnjl2} to include the Polyakov loop 
effective potential \cite{polyd1,polyd2,pisarski1,pisarski2}. While the 
NJL part is supposed to give the correct chiral properties, the 
Polyakov loop part should simulate the deconfinement physics. Indeed 
such a synthesis worked well to firstly confirm the "coincidence" of 
onset of chiral restoration and deconfinement as observed in Lattice 
simulations (see discussions in Ref.\cite{digal}) in the PNJL model 
\cite{pnjl1}. Secondly, the pressure, scaled pressure difference, number
density and the interaction measure were computed in Ref.\cite{pnjl2} 
for two quark flavours, and all the quantities compared well with the 
lattice data. This model is supposed to work upto an upper limit of 
temperature, because here the gluon physics is contained only in a 
static background field that comes in the Polyakov loop. However, 
transverse degrees of freedom will be important for $T > 2.5 T_c$ 
\cite{tranv}.

For the details of the PNJL model parameterization, the reader is 
referred to Ref.\cite{pnjl2}. Our starting point is the thermodynamic 
potential per unit volume given by,

\beqa
\Omega&=&{\cal U}\left(\Phi,\bar{\Phi},T\right)+\frac{\sigma^2}{2G}-
2N_f\,T\int\frac{\mathrm{d}^3p}{\left(2\pi\right)^3}
\left\{ \ln\left[1+3\left(\Phi+\bar{\Phi}\mathrm{e}^
{-\left(E_p-\mu_0\right)/T}\right)\mathrm{e}^{-\left(E_p-\mu_0\right)/T}
 + \mathrm{e}^{-3\left(E_p-\mu_0\right)/T}\right]\right . \nonumber\\
&+& 
\left . 
\ln\left[1+3\left(\bar{\Phi}+\Phi\mathrm{e}^{-\left(E_p+\mu_0\right)/T}
\right)\mathrm{e}^{-\left(E_p+\mu_0\right)/T}+
\mathrm{e}^{-3\left(E_p+\mu_0\right)/T}\right] \right\}
- 6N_f\int\frac{\mathrm{d}^3p}{\left(2\pi\right)^3}{E_p}
\theta\left(\Lambda^2-\vec{p}^{~2}\right) ~~~.
\label{omega}
\eeqa

Here, ${\cal U}\left(\Phi,\bar{\Phi},T\right)$ is the effective
potential for the traced Polyakov loop $\Phi$ and its conjugate
$\bar{\Phi}$, and $T$ is the temperature. The functional form of
the potential is,

\beqa
{\mathcal{U}\left(\Phi,\bar{\Phi},T\right)\over T^4} =
-{b_2\left(T\right)\over 2 }\bar{\Phi} \Phi-
{b_3\over 6}\left(\Phi^3+
{\bar{\Phi}}^3\right)+ {b_4\over 4}\left(\bar{\Phi} \Phi\right)^2 ~~~,
\label{uu}
\eeqa

with

\beqa
b_2\left(T\right)=a_0+a_1\left(\frac{T_0}{T}\right)
+a_2\left(\frac{T_0}{T} \right)^2+a_3\left(\frac{T_0}{T}\right)^3~~~.
\label{bb}
\eeqa

The coefficients $a_i$ and $b_i$ were fitted from Lattice data of
pure gauge theory. The parameter $T_0$ is precisely the transition
temperature for this theory, and as indicated by Lattice data its
value was chosen to be 270 MeV \cite{tcpg1,tcpg2,tcpg3}. With the 
coupling to NJL model the transition doesn't remain first order. 
In this case from the peak in $d\Phi/dT$ the transition (or crossover) 
temperature $T_c$ comes around 230 MeV. The authors in Ref.\cite{pnjl2} 
then reduce the $T_0$ to 190 MeV such that the $T_c$ becomes about 180
MeV, commensurate with Lattice data with two flavours of dynamical 
fermions \cite{tcdyn}. However, we shall keep using $T_0=\rm{270~MeV}$,
since for $T_0=\rm{190~MeV}$, there is about 25 MeV shift in the chiral
and deconfinement transitions with all other model parameters remaining
fixed, as compared to less than 5 MeV shift for $T_0=\rm{270~MeV}$. We
have checked that for both the values of $T_0$, the susceptibilities 
we measure, when plotted against $T/T_c$ show very little dependence 
on $T_0$.

The other notations in (\ref{omega}) are as follows. $\sigma$ is
the auxiliary field introduced via bosonization techniques in the
PNJL Lagrangian. $<\sigma>=G<\bar{\psi}\psi>$ is the chiral condensate.
$G$ is the effective coupling strength of a local, chiral symmetric
four-point interaction. $N_f$ denotes number of flavours. 
$E_p=\sqrt{\vec{p}^{~2}+m^2}$, where 
$m=m_0-\langle\sigma\rangle=m_0-G\langle\bar{\psi}\psi\rangle$, with
$m_0=m_u=m_d$ being the value of current quark mass. $\Lambda$ is the
3-momentum cutoff in the NJL model. $\mu_0$ is the quark number chemical
potential.

Given the thermodynamic potential, our job is to minimize it with 
respect to the fields $\sigma$, $\Phi$ and $\bar{\Phi}$, and calculate
the necessary quantities with the values of the fields so obtained.

\section{Formalism}

Here we give the details of the methods by which we have obtained the
various thermodynamic quantities from the PNJL model. 

\subsection{Taylor expansion of Pressure}

The pressure as a function of temperature $T$ and baryon chemical 
potential $\mu_0$ is given by,

\beqa
P(T,\mu_0) = -\Omega(T,\mu_0) ~~~.
\label{prsu}
\eeqa

The first derivative of pressure with respect to $\mu_0$ gives the quark
number density. The second derivative is the quark number susceptibility
(QNS), which should show a power law divergence close to the CEP.
On the Lattice where direct simulation for non-zero $\mu_0$ is not 
possible, the QNS and higher order derivatives (HODs) computed at 
$\mu_0=0$ are used as Taylor expansion coefficients to extract chemical 
potential dependence of pressure. In fact the convergence of such an 
expansion has been tested with the HODs to obtain the CEP. At this point
we should mention that in the present model the isospin chemical 
potential $\mu_I$ has not been included, which will be necessary to 
study the isospin asymmetric QCD matter. Also, the presence of diquark 
physics needs to be incorporated in the model to have a complete study 
of PNJL model for large $\mu_0$.

So, given $\Omega$ as in the PNJL model we first solve  numerically
for $\sigma$, $\Phi$ and $\bar{\Phi}$ using the following set of 
equations,

\beqa
{\partial\Omega\over\partial\sigma} = 0 ~,~~~~~
{\partial\Omega\over\partial\Phi} = 0 ~,~~~~~
{\partial\Omega\over\partial \bar{\Phi}} = 0 ~.
\label{solv}
\eeqa

The values of the fields so obtained can then be used to evaluate all 
the thermodynamic quantities in mean-field approximation. To obtain the
transition point we need to look at the behaviour of the temperature 
derivatives of the fields. The method used here to obtain the 
quantities $\partial \Phi/\partial T$ and $\partial \sigma/\partial T$
is as follows. First we numerically solve Eqn. (\ref{solv}) for each 
value of temperature $T$ for $\mu_0=0$. The temperature difference 
between consecutive data is 0.1 MeV. Then we obtain the slope of the 
($T,\Phi(T)$) and the ($T,\sigma(T)$) curves. However, instead of using
the difference method of extracting the slope, we do a fit to a Taylor
expansion of the fields as a function of $T$, around the point where 
the slopes of the fields are required. We used a quadratic fitting
function and the first order coefficient then gives us the required slope.
We did the analysis more carefully near the transition where the slopes
were obtained for every 1 MeV difference. Our results are identical to
that obtained in Ref.\cite{pnjl2}.

The field values obtain from Eqn.(\ref{solv}) are then put back into 
$\Omega$ to obtain pressure from (\ref{prsu}). We can then expand the 
scaled pressure as,

\beqa
{P(T,\mu_0) \over T^4}= \sum^{\infty}_{n=0}c_n(T)\left({\mu_0 \over
T }\right)^n ~~~,
\label{tay}
\eeqa

where,

\beqa
c_n(T) = {1 \over n!} {\partial^n \left ({P(T,\mu_0) / T^4} \right ) 
         \over \partial \left({\mu_0 \over T }\right)^n}\Big|_{\mu_0=0} ~~~.
\eeqa

We shall use the expansion around $\mu_0 = 0$. In this expansion, 
the odd terms vanish due to CP symmetry. We extract the expansion
coefficients upto eight order. This has been motivated by the fact
that on the Lattice the most recent results are also obtained upto
this order. 

In general, to obtain the Taylor coefficients of pressure, one can 
use either of the two methods:

\noindent
\underline{Method $(a)$:}
First the pressure is obtained as a function of $\mu_0$ for each value 
of $T$, and then fitted to a polynomial in $\mu_0$. The quark number 
susceptibility (QNS) and all other higher order derivatives (HODs) are 
then obtained from the coefficients of the polynomial extracted from
the fit.

\noindent
\underline{Method $(b)$:}
First obtaining the expressions for the derivatives of the pressure 
with respect to $\mu_0$ for the Taylor coefficients and then use the 
values of $\sigma$, $\Phi$ and $\bar{\Phi}$ at zero chemical potential
into these expressions. 

In any exact computation, these two methods should yield identical 
results. The Lattice however at present cannot use method $(a)$ due 
to the complex determinant problem. On the other hand since we are
using the mean field analysis, method $(b)$ would give us wrong
results as the mean fields used would be insensitive to $\mu_0$.
In this work we have computed all the observables using method $(a)$.
We have expanded the pressure in an eighth order polynomial in
$\mu_0$ with even terms only (the odd terms should be zero due to
CP symmetry).

\subsection{$C_V$, $v_s^2$ and $\cC$}

Given the thermodynamic potential $\Omega$, the energy density 
$\epsilon$ is obtained from the relation,

\beqa
\epsilon 
   = - T^2 \left . {\partial (\Omega/T) \over \partial T} \right |_V
   = - T \left . {\partial \Omega \over \partial T} \right |_V 
              + \Omega ~~~.
\eeqa

The rate of change of energy density $\epsilon$ with temperature 
at constant volume is the specific heat $C_V$ which is given as,

\beqa
C_V = \left . {\partial \epsilon \over \partial T} \right |_V 
    = - \left . T {\partial^2 \Omega \over \partial T^2} \right |_V ~~~.
\label{sph}
\eeqa

For a continuous phase transition one expects a divergence in $C_V$,
which, as discussed earlier, will translate into highly enhanced 
transverse momentum fluctuations or highly suppressed temperature
fluctuations if the dynamics in relativistic heavy-ion collisions is
such that the system passes close to the CEP. 

The square of velocity of sound at constant entropy $S$ is given by,

\beqa
v_s^2 = \left . {\partial P \over \partial \epsilon} \right |_S 
      = \left . {\partial P \over \partial T} \right |_V \left /
        \left . {\partial \epsilon \over \partial T} \right |_V \right .
      = \left . {\partial \Omega \over \partial T} \right |_V \left /
        \left . T {\partial^2 \Omega \over \partial T^2} \right |_V 
        \right .  ~~~.
\label{sps}
\eeqa

Since the denominator is nothing but the $C_V$, a divergence in specific
heat would mean the velocity of sound going to zero at the CEP. 

The conformal measure $\cC$ is given by,

\beqa
\cC = {\Delta \over \epsilon} = {\epsilon - 3P \over \epsilon} 
    \simeq 1 - 3v_s^2 ~~~.
\label{cnf}
\eeqa

Thus, a minima in the velocity of sound as expected near a phase
transition or crossover may translate to a maxima of $\cC$. Considering
the last relation we see that at asymptotic temperatures where the 
$v_s^2$ goes to the ideal gas value of 1/3, the conformal measure should
go to the conformal limit $\cC = 0$.

Given the relations (\ref{sph}), (\ref{sps}) and (\ref{cnf}) we perform
the same exercise of obtaining the $\Omega(T,\mu_0=0)$ from the PNJL 
model. We then obtain the Taylor expansion coefficient of 
$\Omega(T,\mu_0=0)$ around the temperatures at which we wish to obtain 
the values of $C_V$, $v_s^2$ and $\cC$. The expansion is done upto 
second order and fitted with the numerical values. 
From the first and second coefficients of this expansion we obtain 
$\partial \Omega / \partial T$ and $\partial^2 \Omega/ \partial T^2$. 
These are then put into the corresponding expressions for the 
thermodynamic quantities.

\section{Results}

\begin{figure}[!tbh]
   \includegraphics{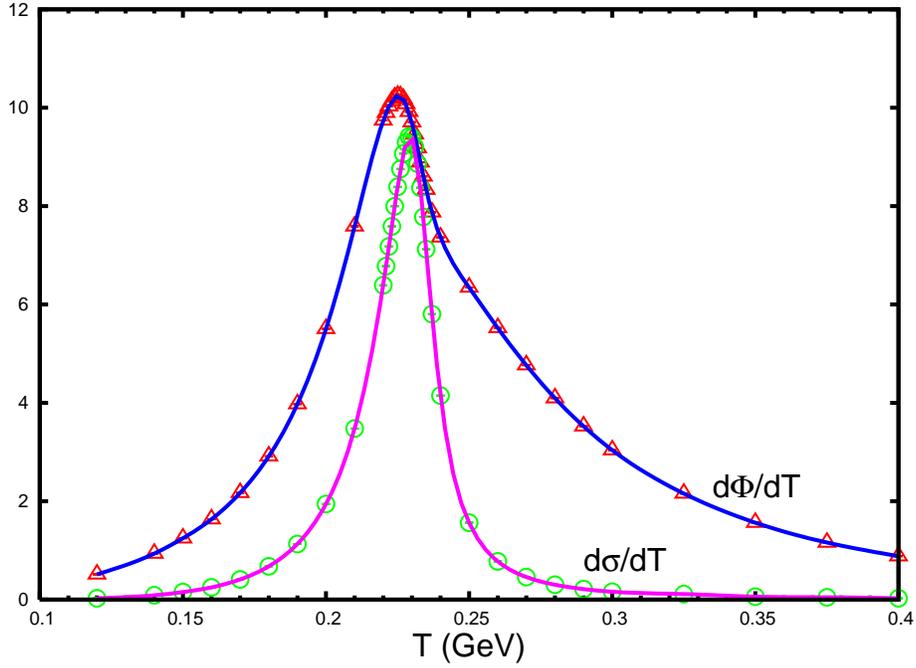}
   \caption{$\partial \Phi/\partial T$ and $\partial \sigma/\partial T$
	    as a function of temperature T (GeV) (also obtained in 
            \cite{pnjl2} Fig.4b).The symbols show our data, the lines 
            are spline fitting.
      }
\label{fg.dphdsg}\end{figure}

\begin{table}[!htbp]
  \begin{center}\begin{tabular}{|c|c|c|l|l|l|c|}  \hline
  $T$  &Range of     &Order of  & $c_0$ & $c_2$ & $c_4$ & Reduced\\
  (MeV)&$\mu_0$ (MeV)&Polynomial&       &       &       & $\chi^2$\\
  \hline
  120  & 0 -- 200    & 8        &0.0050743(8)&0.001450(6)&0.00125(1)&2.08836e-10\\ 
       &             & 4        &0.00540(1)  &-0.00080(3)&0.00325(1)&1.0377e-07\\
       & 0 -- 100    & 8        &0.005069(1) &0.00162(3) &0.0004(2) &1.96592e-10\\
       &             & 4        &0.0050741(9)&0.001422(8)&0.00145(1)&2.12054e-10\\
  \hline 
  180  & 0 -- 170    & 8        &0.0691975(2)&0.032304(4)&0.02599(2)&9.14908e-12\\
       &             & 4        &0.069407(9) &0.02766(6) &0.03972(8)&4.05665e-08\\
       & 0 -- 100    & 8        &0.0691960(2)&0.03240(1) &0.0251(2) &7.32281e-12\\
       &             & 4        &0.0692026(4)&0.031951(8)&0.02934(3)&4.2536e-11\\
  \hline
  227  & 0 -- 100    & 8        &0.3890060(1)&0.317329(9)&0.2775(2) &1.22448e-12\\
($T_c$)&             & 4        &0.3889960(7)&0.31812(2) &0.2730(1) &1.39197e-10\\ 
       & 0 -- 50     & 8        &0.3890060(1)&0.31732(5) &0.278(5)  &1.22912e-12\\
       &             & 4        &0.3890060(1)&0.31731(1) &0.2802(3) &1.22551e-12\\ 
  \hline
  350  & 0 -- 200    & 8        &2.380420(0) &0.7585680(6)&0.103069(9)&3.67489e-14\\
       &             & 4        &2.380410(0) &0.759109(6) &0.09791(2) &5.0959e-11\\
       & 0 -- 100    & 8        &2.380420(0) &0.758570(4) &0.1029(2)  &3.77728e-14\\
       &             & 4        &2.380420(0) &0.758605(1) &0.10169(1) &5.44908e-14\\
  \hline
  450  & 0 -- 200    & 8        &3.120020(0) &0.8244970(4)&0.094363(10)&4.99485e-15\\
       &             & 4        &3.120010(0) &0.824656(2) &0.091886(10)&1.68283e-12\\
       & 0 -- 100    & 8        &3.120020(0) &0.824497(2) &0.0944(2)   &5.02939e-15\\
       &             & 4        &3.120020(0) &0.8245070(5)&0.09372(1)  &5.36979e-15\\
  \hline
  \end{tabular}\end{center}
  \caption[dummy]{Representative values of fitted Taylor coefficients
to show the quality of fit. For each temperature, data in first line is
used for the figures. (Here $\chi^2$ is same as least square.)}
\label{tb.fit}\end{table}

As mentioned earlier we use the parameterization of the PNJL model 
as given in Ref.\cite{pnjl2} and reproduced the behaviour of 
$\partial\Phi/\partial T$ and $\partial\sigma/\partial T$ as shown
in Fig.\ref{fg.dphdsg}. For $\mu_0=0$, the transition is 
quite sharp at $T_{chiral} \simeq \rm{229~MeV}$ and 
$T_{deconfinement} \simeq \rm{225~MeV}$. 
For our subsequent results we shall present the 
quantities as a function of temperature in units of a cross-over 
temperature which is taken to be $T_c= \rm{227~MeV}$. In view of the
limitations of the model at high temperatures as mentioned earlier, we
obtain results for thermodynamic quantities upto about $2.5 T_c$.

\subsection{Taylor expansion of Pressure}
\begin{figure}[!tbh]
   \includegraphics{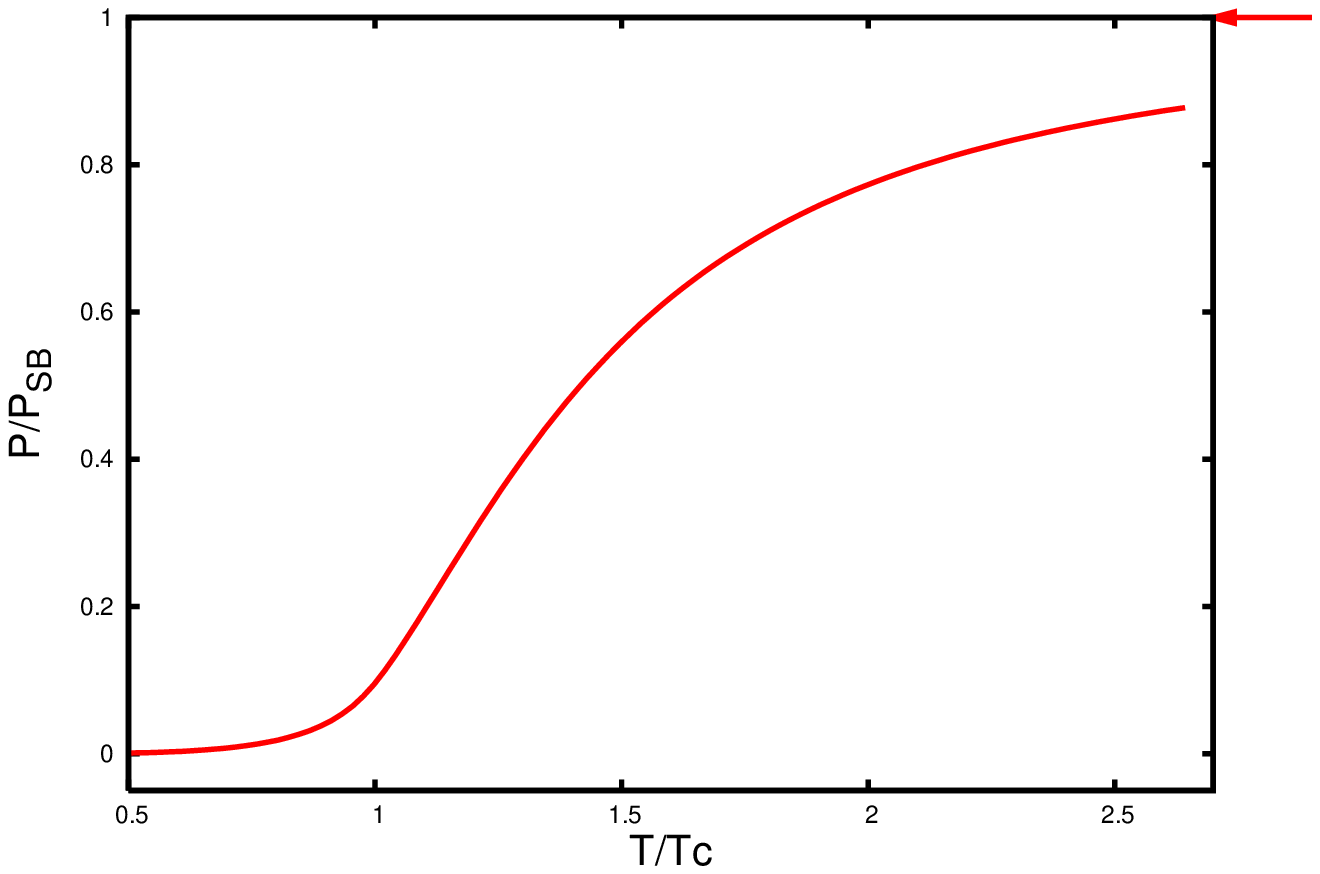}
   \caption{Pressure scaled with Stefan-Boltzmann pressure as function
            of temperature for $\mu_0=0$ from PNJL model. Arrow
            on the right indicates ideal gas value.
	   }
\label{fg.press}\end{figure}

\begin{figure}[!tbh]
   \includegraphics{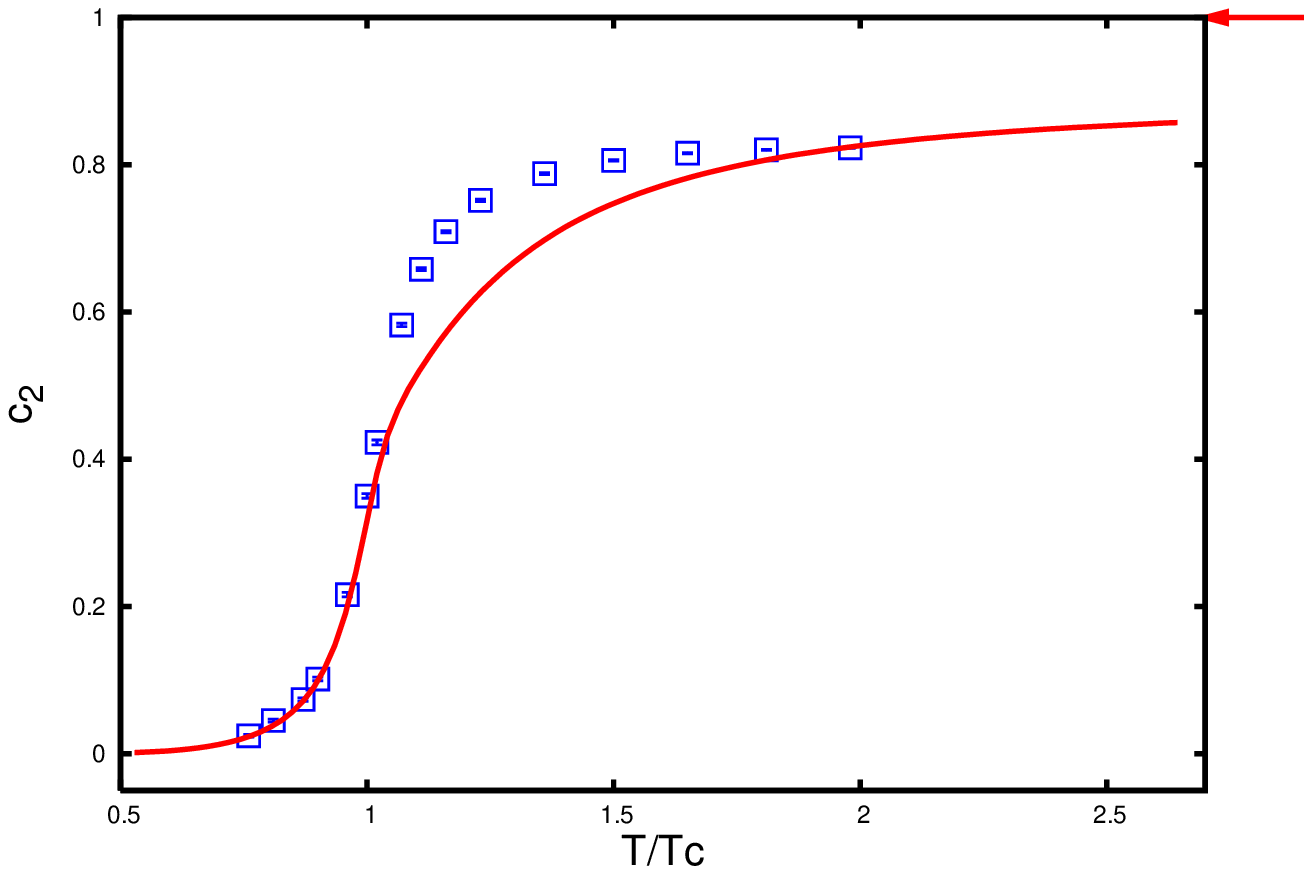}
   \caption{The QNS as a function of $T/T_c$. Symbols are Lattice data 
            \cite{sixx}. Arrow on the right indicates ideal gas value.
	   }
\label{fg.scnord}\end{figure}

We now present the pressure and its derivatives with respect to $\mu_0$
at $\mu_0=0$. The pressure was fitted to a polynomial in $\mu_0$ using
the GNU plot program at different values of temperature. The maximum 
range of $\mu_0$ was chosen to be 200 MeV. However near $T_c$ the 
$\chi^2$ (which in this case is same as the least square) of the fit 
varies rapidly with the variation of range of $\mu_0$ over which the 
fit was done, and the actual range was chosen from minima of 
$\chi^2$. The data points were spaced by 0.1 MeV. 

In table \ref{tb.fit} we present the fitted values of 
the Taylor coefficients for a few values of temperature to show the
dependence of these fitted coefficients on the range of $\mu_0$ and
number of terms in the polynomial. Due to limitations of numerical
accuracy and time, we chose to take a maximum of only upto the eighth 
order term in $\mu_0$. We hope to obtain higher order coefficients
in future. The coefficients seem to be quite robust. For each 
temperature in the table the coefficients in the first case was 
actually used in the final result.

Fig.\ref{fg.press} displays the pressure scaled with that
of Stefan-Boltzmann (SB) gas ($P/P_{SB}$) as a function of $T/T_c$. Upto
$2.5 T_c$ the pressure grows from almost zero at low temperatures to 
about 90\% of its ideal gas value. This is a bit high when compared to 
continuum estimates on the Lattice \cite{latpr}, which is about 80\%
of $P_{SB}$. A comparison of our results with the PNJL result in Fig.7a
of Ref.\cite{pnjl2}, shows a near perfect match, though 
$T_0= \rm{270~MeV}$ in our measurements compared to their value of 
$T_0=\rm{190~MeV}$. 

\begin{figure}[!tbh]
   \includegraphics{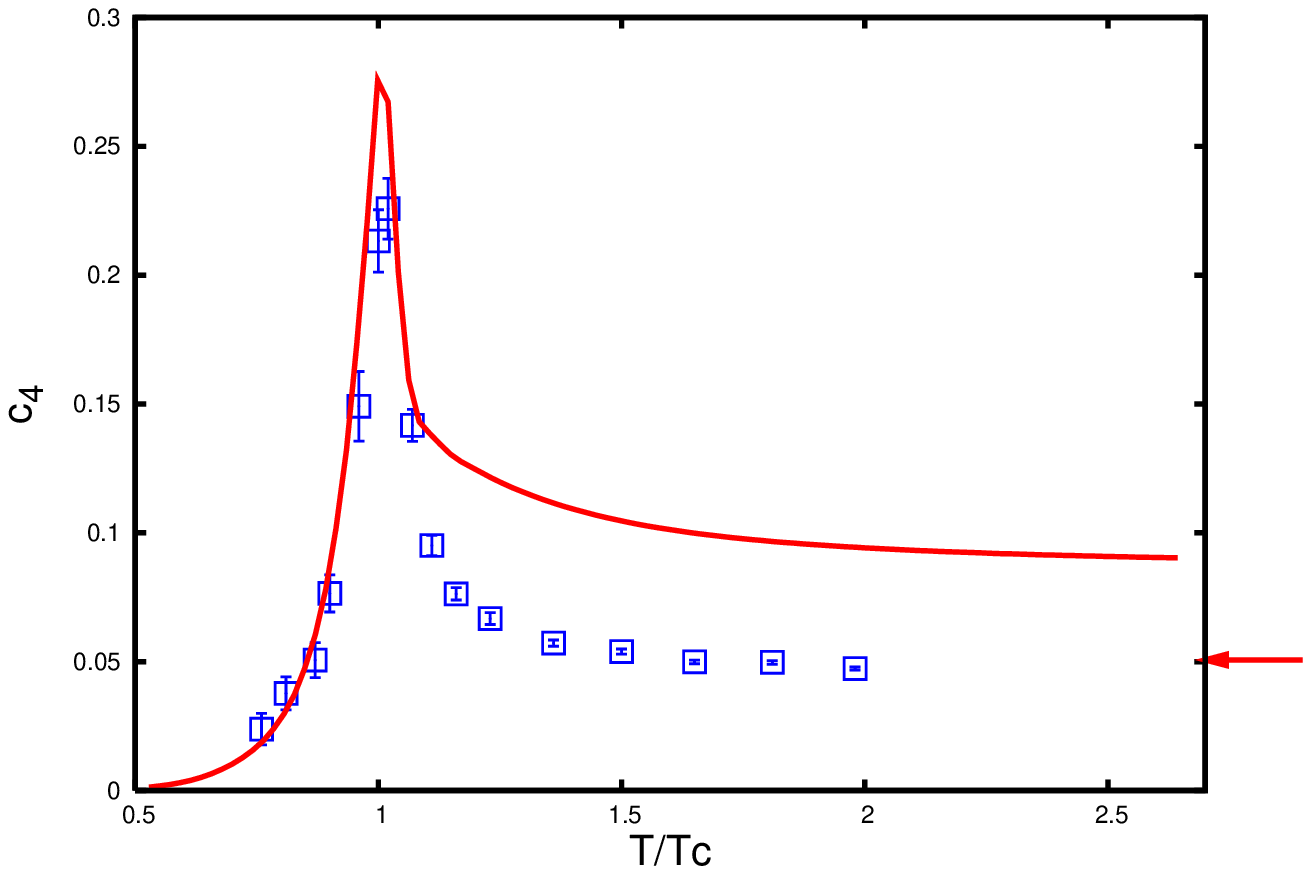}
   \caption{The $c_4$ as a function of $T/T_c$. The solid line is from
            PNJL model, Symbols are Lattice data \cite{sixx}. Arrow
            on the right indicates ideal gas value.
	   }
\label{fg.hrord4}\end{figure}

Next we show the comparison of the coefficients $c_2$, $c_4$ and $c_6$ 
obtained by us with Lattice data available in Table 3.2 of Ref.\cite{sixx}. 
Fig.\ref{fg.scnord} shows the variation of the QNS $c_2$ with $T/T_c$.
This shows an order parameter-like behaviour. Similar behaviour has
been observed in another model study \cite{sanjay} using the Density
Dependent Quark Mass (DDQM) model. At higher temperatures the $c_2$
reaches almost 85 \% of its ideal gas value, consistent with Lattice
data. 

\begin{figure}[!tbh]
   \includegraphics{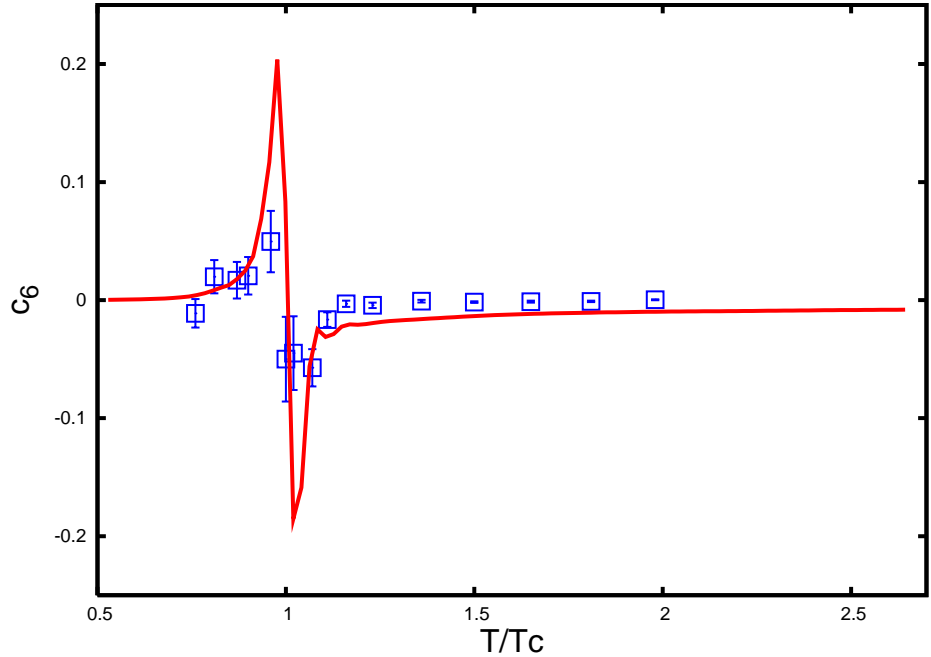}
   \caption{The $c_6$ as function of $T/T_c$. The solid line is from
            PNJL model, Symbols are Lattice data \cite{sixx}.
	   }
\label{fg.hrord6}\end{figure}

\begin{figure}[!tbh]
   \includegraphics{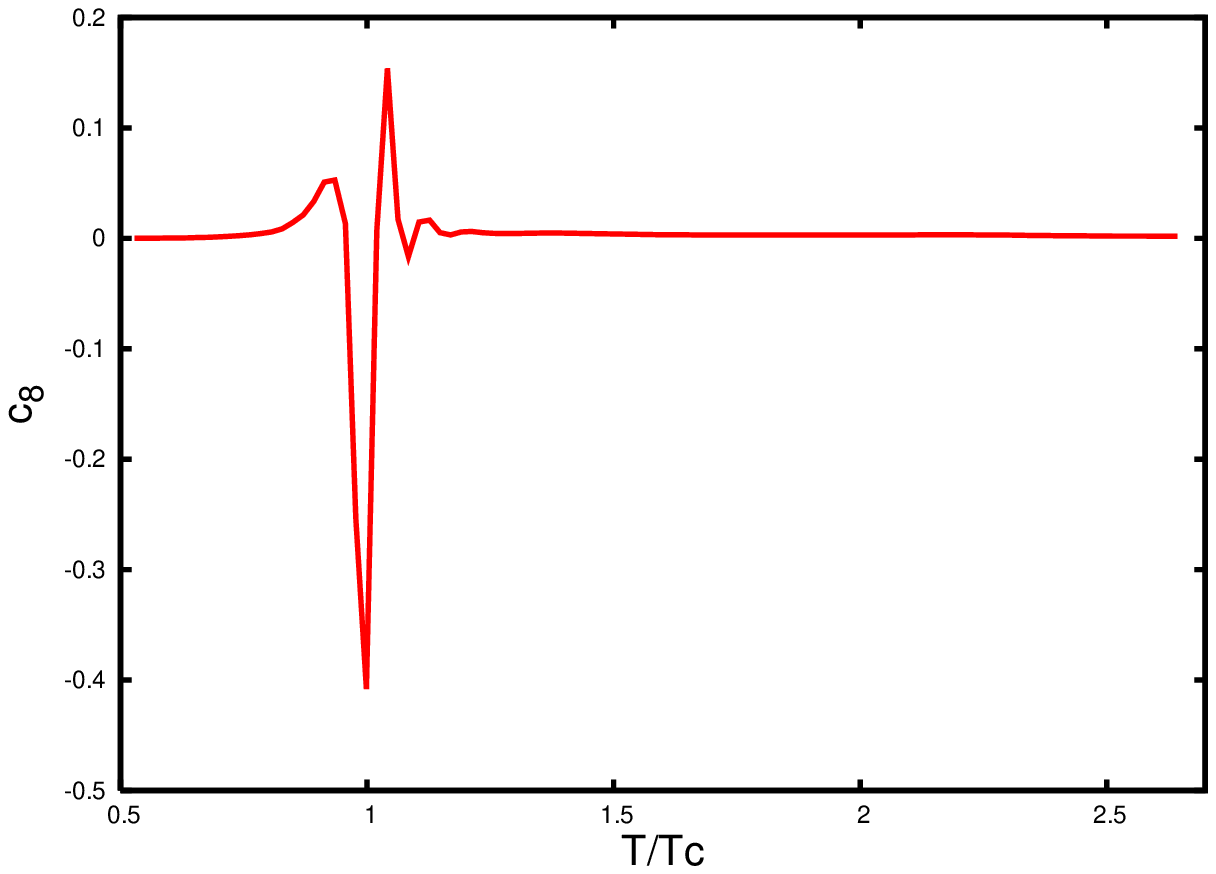}
   \caption{The $c_8$ as function of $T/T_c$ from PNJL model.
	   }
\label{fg.hrord8}\end{figure}

The fourth order derivative $c_4$, which can then be thought of as 
the "susceptibility" of $c_2$ shows a peak at $T=T_c$ (Fig.\ref{fg.hrord4}). 
Near the transition temperature $T_c$, the effective model should work 
well and we observe that the structure of $c_4$ is quite consistent with 
present day Lattice data \cite{eight,sixx}. Just above $T_c$ however, 
there is a significant difference between our results of $c_4$ and that of 
Ref.\cite{sixx}. While the Lattice values converge to the SB limit, ours
is almost double of that value and shows only a weak convergence towards
the SB limit. However, our results are consistent with the values of 
$\chi_{40}$ in Fig.3 in Ref.\cite{eight}, though it has been argued 
\cite{smalmu4} that these data would come down to the SB limit once a 
correct continuum limit is taken. Note that in the SB limit both $c_2$ 
and $c_4$ have only fermionic contributions. We expect that because the
coupling strength is still large in this temperature regime it is 
unlikely that $c_4$ should go to the SB limit within $T < 2.5 T_c$. 
Moreover, the quark masses used in Ref.\cite{sixx} is considerably large
($m/T = 0.4$) to expect fermionic observables to go to the SB limit.
However, it is possible that our overestimation is due to the use of
mean field approximation. Recently, in a quasi-particle model \cite{soff} the 
Taylor coefficients have been found to match the Lattice data quite 
accurately. This model uses $c_2$ to fix the $T$ and $\mu$ dependence
of an effective coupling in the self energy and then predicts the
values of the higher order coefficients. In contrast, the PNJL model
uses only pure gauge results on the lattice to fix the parameters of
the Polyakov loop effective potential and then predicts all the
coefficients including $c_2$.

The HODs $c_6$ and $c_8$ are plotted in Fig.\ref{fg.hrord6} and 
Fig.\ref{fg.hrord8} respectively. At high temperatures both of these 
HODs converge to zero. The structure near $T_c$ is more interesting.
Compared to other models the pattern of the HODs show much better 
resemblance to the Lattice results. The hadron resonance gas model 
describes the Lattice data well below $T_c$, but fails for $T > T_c$ 
\cite{hdrs}. The recently proposed \cite{shubd} scenario of coloured 
bound states also compares the HODs from Lattice data. However, their
comparison at the present state is still not completely satisfactory.

\subsection{$C_V$, $v_s^2$ and $\cC$}

\begin{figure}[!tbh]
   \includegraphics{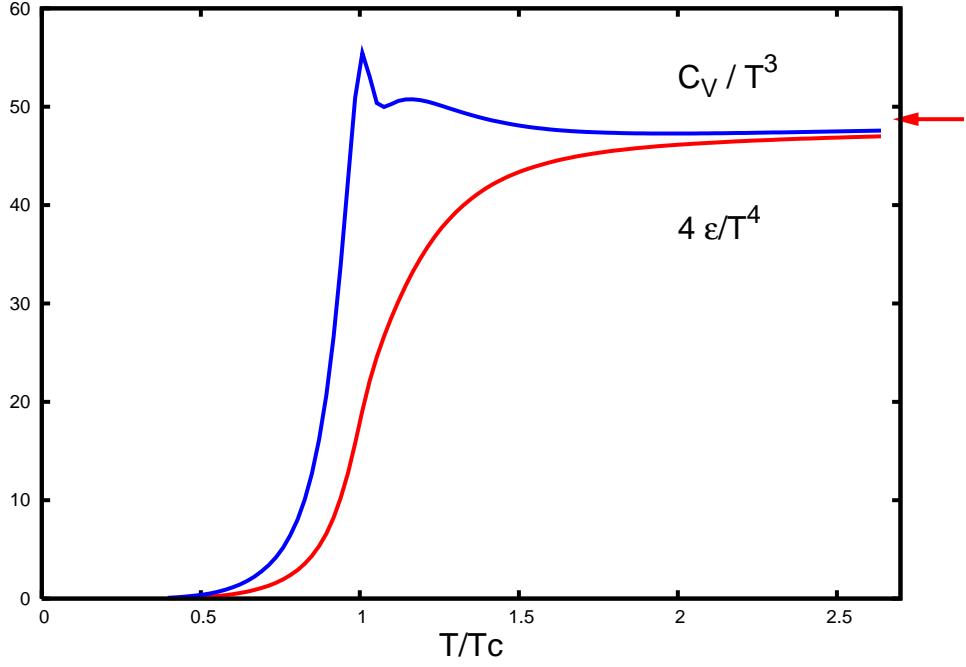}
   \caption{$C_V/T^3$ and $4 \epsilon/T^4$ as function of $T/T_c$.
           The arrow on the right shows the ideal gas value.
           }
\label{fg.cv}\end{figure}

We now discuss the results for the other thermodynamic quantities
that we have computed. First we consider the specific heat. As shown 
in Fig.\ref{fg.cv}, $C_V$ grows with increasing temperature and reaches
a peak at $T_c$. Then it decreases sharply for a short range of 
temperature. Thereafter it shows a broad but shallow bump around 
$T=\rm{270~MeV}$ and then gradually goes to a value little below the 
ideal gas value. Comparing with the recent lattice data for pure glue 
theory (Fig.3f, Ref.\cite{swa2}), we see that the difference from the
ideal gas value is far less in our case. For comparison, we have 
also plotted the values of $4 \epsilon/T^4$, at which the specific 
heat is expected to coincide for a conformal gas. As is clear from 
the plot, the values almost coincide at the large temperatures but
not as perfectly as in Ref.\cite{swa2}. 

\begin{figure}[!tbh]
   \includegraphics{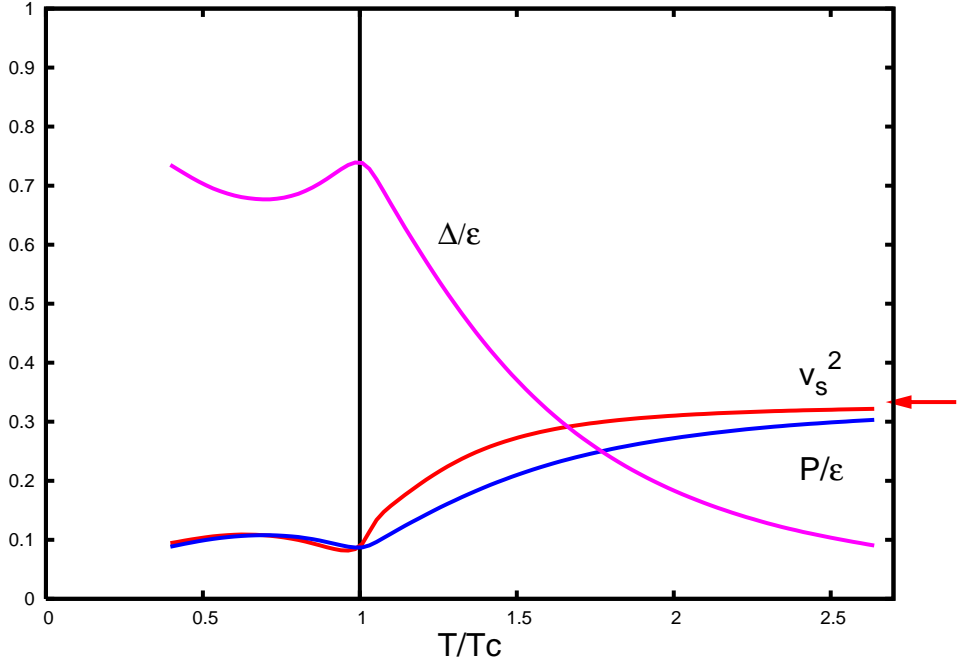}
   \caption{Squared velocity of sound $v_s^2$ and conformal measure
           $\cC=\Delta/\epsilon$ as function of $T/T_c$. The arrow on
           the right shows the ideal gas value for $v_s^2$. For 
           comparison with $v_s^2$ we also plot the ratio $P/\epsilon$.
           }
\label{fg.ccs}\end{figure}

Let us now consider the speed of sound and the conformal measure. In 
view of the possible relation between the two as indicated in 
(\ref{cnf}), we plot both these quantities together in Fig.\ref{fg.ccs}. 
As can be seen from the figure, the two quantities indeed behave in
such a correlated manner. We note here that $\cC$ has been computed from
the zeroth and first order coefficients of $\Omega(T,\mu_0=0)$, whereas
the $v_s^2$ has been measured from the first and second order 
coefficients. However the relation implied is not exact as can be seen 
in the figure. The value of $P/\epsilon$ matches with that of $v_s^2$ 
for $T<T_c$ and also goes close again above $2.5 T_c$. But in between 
these two limits the $v_s^2$ is distinctly greater than $P/\epsilon$. 
Thus, $\cC$ would go to zero much faster if we replace $P/\epsilon$ by
$v_s^2$ for computing $\cC$. 

The $v_s^2$ is close to its ideal gas value at the temperature of about
$2.5 T_c$. This is close to the results for pure glue theory on the
Lattice as reported in Ref.\cite{swa2} and also that with 2 flavour 
Wilson Fermions in Ref.\cite{khan}, with 2+1 flavours of staggered 
quarks reported in Ref.\cite{szabo}, and with improved 2 flavour 
staggered fermions ($P/\epsilon$ was measured in this case). This 
possibly refers to an intriguing fact that $v_s^2$ is dominated by the 
gluonic degrees of freedom at least for large temperatures. However, 
near $T_c$ the $v_s^2$ in Ref.\cite{swa2} goes to a minimum value above
0.15, whereas we find the minima going close to 0.08, consistent with 
simulations with dynamical quark in Refs.\cite{khan,szabo}, and 
remarkably close to the softest point $P/\epsilon = 0.075$ in 
Ref.\cite{isentropp}. This is expected as the scalar $\sigma$ is 
supposed to be important in chiral dynamics for small quark masses.
Even upto temperatures as low as $0.5 T_c$ the $v_s^2$ in our case 
doesn't go much above 0.1. This is also in contrast with the values 
reported in Ref.\cite{beda}, where a confinement model \cite{wws} has 
been used. These authors find a value of about $v_s^2=0.15$ near $T_c$
and a value of 0.2 near $0.5 T_c$. On the other hand $v_s^2$ 
(isothermal) measured in Ref.\cite{sanjay} using the DDQM model, shows
a better agreement with our results.  

The values for the conformal measure $\cC$ also closely resembles the
Lattice data of Ref.\cite{swa2}. Near $T_c$ there is a slight departure.
We find the value ranging from 0.7 to 0.75, whereas those authors find
it to vary around 0.6 to 0.7 for pure glue theory. However, note that 
the dip at temperatures less than $T_c$ is prominent in both the cases.
At even lower temperatures we find $\cC$ to increase. For a 
non-relativistic ideal gas, the ratio of $P/\epsilon$ should go to zero,
and thus $\cC$ should then go to 1. Thus at lower temperatures such an
increase is indeed expected. On the other hand at high temperatures 
either an ideal gas or a conformal behaviour should be recovered for
which $\cC$ should go to zero. Our results in this respect resemble the
pure gauge Lattice results very closely.

\section{Discussions and Summary}

We have studied several thermodynamic quantities of recent interest in
QCD, in the framework of Polyakov-loop extended NJL (PNJL) model. 
Inclusion of Polyakov loop in the NJL model covers both the confinement 
as well as chiral symmetry breaking; the two most important 
characteristics of low energy QCD. The Polyakov-loop provides with the 
order parameter for deconfinement transition while the quark condensate 
acts as the order parameter for chiral transition.

 It is well known that the measurement of various fluctuations is of
crucial importance to understand the characteristics of transition of
confined matter to deconfined state. So, using the PNJL model, we have
computed the derivatives of pressure upto eight order. It is expected 
that the susceptibilities might provide the most direct evidence for 
the order of QCD phase transition. In fact, a well defined peak in the 
susceptibility can indicate the cross over transition. On the other 
hand, a sharp diverging behaviour would indicate the existence of CEP.

In the present paper, we have kept our computation close to $\mu_0 = 0$
axis. The second derivative of pressure $c_2$ rises steeply near $T_c$
and then saturates near $T = 2 T_c$. This behaviour indicates that near
the transition the quark number fluctuations increase with temperature
and then gradually saturates at higher temperatures. Such a behaviour
is expected both from lattice as well as phenomenological studies. $c_4$
has a peak around $T_c$, whereas, both $c_6$ and $c_8$ show rapid 
variation across $T_c$. Considering the inherent differences of model 
and lattice approaches, our results are in good agreement with the 
lattice data. There is however a significant difference found in the 
values of $c_4$ in the temperature range starting from a little above
$T_c$. We speculate this may have occurred due to the neglect of 
fluctuations around the mean field in our computations; though one
should keep in mind the very large masses used in Lattice computations
in contrast to the value of $5.5 {\rm MeV}$ used in the PNJL model. 

We have also 
calculated the specific heat, speed of sound and the conformal measure.
These results again are in close agreement with Lattice data. There 
seems to be a dominance of gluonic contribution at high temperatures. 
Near $T_c$ however there are small but significant difference in 
Lattice data with pure QCD and that with dynamical quarks. Our results 
in this temperature region match well with the latter set.

There are several issues that need to be addressed for a better 
understanding of the physics of QCD phase transition. NJL model works 
with constant four point coupling strength, the running coupling being
averaged over a limited low energy kinematic domain. To make contact 
with QCD at high temperatures this coupling should have proper 
temperature dependence. Moreover, the parameters in the Polyakov-loop
part should be tuned to get both the deconfinement and chiral transition
within a temperature difference of couple of MeVs at most and also below,
say, 200 MeV, as indicated by the Lattice data. Note that similar
parameter fitting of the Polyakov-loop in the PNJL model led the authors
in Ref.\cite{pisarski2} to couple the Polyakov-loop and 
$SU(3) \times SU(3)$ linear sigma model to study the $K/\pi$ ratio. 
However, there are more predictions in the present model, the most 
significant being the possibility of studying the QCD transition at 
finite baryon densities. Above and all one should also worry about the 
mean-field approximation involved here. For the Polyakov loop part,
it would be worthwhile to explore approches similar to 
Ref.\cite{enrique}, where a local Polyakov loop has been coupled to 
the chiral Lagrangian.

We have not yet studied the system with large chemical potentials.
As already noted in Ref.\cite{pnjl2}, the PNJL model does have the 
requisite physics of a first order transition for low temperatures 
and high chemical potential, a cross-over for $\mu_0=0$ and thus a CEP
for some $T=T_E$ and $\mu_0=\mu_{0_E}$. The details in this case
may be compared to other NJL model studies with high $\mu_0$. For
example the chiral symmetry broken phase and normal quark matter 
phase of Ref.\cite{njlpd} show a similar value of the CEP. Moreover it 
has been shown in Ref.\cite{njltype} that with increase of strength 
of the diquark coupling the CEP moves towards higher temperatures and
slightly lower values of $\mu_0$. Hence, the inclusion of diquark
physics may be important to assess the exact location of the phase 
boundary as well as the CEP. 

A natural extension of our work would be to include the isospin
chemical potential and study the behaviour for isospin and electric 
charge fluctuations. We hope to address some of these issues in future.

\section*{Acknowledgments}

We would like to thank the organizers of WHEPP-9 at Institute of
Physics Bhubaneswar, for providing a stimulating environment for 
discussions from where this project originated. We acknowledge useful
discussions with R.V. Gavai. R.R. would like to thank S. Digal, 
S. Gupta, S.J. Hands and S. Mukherjee for useful comments.

\end{document}